\documentclass[aps,prl,superscriptaddress,twocolumn,showpacs,floatfix]{revtex4}
\usepackage{epsfig}
\newcommand{\Trn}{\mbox{Tr}_{\mbox{n}}}
\newcommand{\Sia}{S_{i}^{\alpha}}
\newcommand{\Sja}{S_{j}^{\alpha}}

\newcommand{\Sib}{S_{i}^{\beta}}

\newcommand{\ie}{{\it i.e.\ }}
\begin{document}
\title{Partitioning and modularity of graphs with arbitrary degree distribution}

\author{J\"org Reichardt\footnote{Present Address: Institute for Theoretical Physics, University of W\"urzburg, Am Hubland, 97074 W\"urzburg, Germany}}
\author{Stefan Bornholdt}
\affiliation{Institute for Theoretical Physics, University of Bremen, 
D-28359 Bremen, Germany}
\date{\today}

\begin{abstract}
\noindent We solve the graph bi-partitioning problem in dense graphs with arbitrary degree distribution using the replica method. We find the cut-size to scale universally with $\langle \sqrt{k} \rangle$. 
In contrast, earlier results studying the problem in graphs with a Poissonian degree distribution had found a scaling with $\sqrt{\langle k\rangle}$ [Fu and Anderson, J. Phys. A: Math. Gen. \textbf{19}, 1986]. The new results also generalize to the problem of q-partitioning. They can be used to find the expected modularity $Q$ [Newman and Grivan, Phys. Rev. E, \textbf{69}, 2004] of random graphs and allow for the assessment of statistical significance of the output of community detection algorithms. 
 \end{abstract}
\pacs{89.75.Hc,89.65.-s,05.50.+q,64.60.Cn} \maketitle 

Given a graph or network $\mathcal{G}(N,M)$ of $N$ nodes and $M$ edges, the problem of graph partitioning is finding a partition of the nodes into $q$ equal sized parts, such that the number of edges connecting different parts, the cut-size, is minimal. The solution of this problem has many important practical applications in multiprocessor scheduling for parallel computing, VLSI chip design and data mining \cite{HyperCluster}. 

Due to the problem being NP-complete \cite{Garey}, only heuristic algorithms exist to approximate a solution. Furthermore, one cannot determine in polynomial time, whether a solution found is indeed optimal. The only way to assess the quality of a heuristic algorithm or a particular result is to compare with rigorous bounds on the cut-size. 

In the past, such bounds have been established for Erd\H{o}s-R\'{e}nyi (ER) random graphs \cite{Erdos,Bollobas}  exhibiting Poissonian degree distribution \cite{FuAnd,KanterSompQPart,LaiGold,LiaoRev,LiaoPRL,GoldDom90} or Bethe-Lattices with fixed valence \cite{Wong,Mezard87,Banavar,Oliveira89,SherringtonBethe}, only.  All of these bounds express the expected cut-size in terms of the mean degree of the network. However, this is inappropriate for a large number of problem instances which show fat tailed degree distributions \cite{ClusterChallenges,VLSIPowerLaw} and we will show that the traditional bounds are inappropriate in these cases. 

Recently, the problem of graph partitioning has received considerable 
interest again in a more relaxed version, known as community detection or graph clustering which also has a broad range of applications for data analysis in the life sciences \cite{GuimeraMeta,Palla05} to social science \cite{Arenas}. Given a graph the problem is now finding a partition into $q$ arbitrarily sized communities or clusters which are densely connected internally, but only sparsely connected among each other \cite{Girvan,GuileraReview}. The hypothesis is that in many real world networks, nodes of differing types exist and that nodes of the same type have a higher probability of being linked than nodes of different types \cite{NewmanMixing}. Community detection thus aims at discovering precisely these latent classes of nodes. 

A widely used quality measure for the partition into communities is the so-called 
``modularity'' $Q$ defined by Newman and Girvan \cite{Girvan03}. 
For a partition into $q$ communities of arbitrary size it is defined 
as $Q_q=\sum_{s=1}^q(e_{ss}-a_s^2)$, with $a_s=\sum_{r=1}^qe_{rs}$.
Here, $e_{rs}$ is the fraction of edges falling between nodes in groups $r$ and $s$. Obviously, the larger the distortion between internal edges $e_{ss}$ and expected internal edges $a_s^2$, \ie the more modular a network is, the higher $Q$. By design, $Q<1$. 

In case of $q$ equal sized parts, we have a relation between the cut-size $\mathcal{C}_q$ of the partition and its modularity:
\begin{equation}
\mathcal{C}_q=M\left(\frac{q-1}{q}-Q_q\right).
\label{Cutsize}
\end{equation}
We see that $Q_q$ measures the improvement of the partition over a random assignment of nodes into parts. 

One problem of community detection and all clustering techniques in general is that it is not clear \textit{a priori}, whether latent classes of nodes in a particular network do exist.  Numerical experiments have shown that random networks can be partitioned in a way which has a very high value of $Q$, but this does of course not imply that latent classes of nodes in the network \cite{Guimera,ReichardtPRE} in fact exist. One way to escape this ``deception of randomness'' \cite{EngelBuch} is by comparing the value of $Q$ obtained by clustering the empirical data to an expectation value for $Q$ for an appropriate random null model. 

The partition of a purely random graph with maximum modularity is an equi-partition \cite{ReichardtPRE}. Therefore, solving the problem of graph partitioning for graphs with arbitrary degree distribution, we can also give expectation values for the maximum attainable modularity in purely random graphs and vice versa. These expectation values can be used in the assessment of statistical significance of empirical data analysis. 

The problem of maximizing the modularity is equivalent to finding the ground state of a spin glass \cite{FastGN,ReichardtPRL,ReichardtPRE}. The corresponding Hamiltonian $\mathcal{H}$ for $q=2$ relates to the modularity  $Q_2$ at $\gamma=1$ via :
\begin{equation}
\langle k\rangle N Q_2 = -\mathcal{H} = \sum_{i<j}(A_{ij}-\gamma p_{ij})S_{i}S_{j},
\label{Hamiltonian}
\end{equation}
where $p_{ij}=k_ik_j/\langle k\rangle N$ with $\langle k\rangle$ as the average degree in the network and $A_{ij}$ denotes the adjacency matrix of the graph such that $\sum_jA_{ij}=k_i$, the degree of node $i$. Further, we use Ising spins $S_i=\pm 1$. Following Ref. \cite{ChungLu} $p_{ij}$ may be interpreted as the connection probability in the network due to $\sum_{i<j}p_{ij}=M=\sum_{i<j}A_{ij}$, even though it is not strictly ensured that $p_{ij}<1$ for all $i,j$.
The ground state energy of (\ref{Hamiltonian}) can be written as
\begin{equation}
E_g =  -M+2\mathcal{C}_2+\frac{\gamma}{4M}\left(\sum_{i\in +}k_i-\sum_{i\in -}k_i\right)^2-\frac{\gamma\langle k^2\rangle}{2\langle k\rangle}.
\label{GroundStateCutSize}
\end{equation}
Assuming equal sized parts with the same degree distribution and $\langle k^2\rangle$ being finite, the last two terms in (\ref{GroundStateCutSize}) vanish and we see that (\ref{Cutsize}) holds. Due to the existence of $\langle k^2\rangle$, we can also assume the form of $p_{ij}$ to be a good approximation for the true probability distribution that parameterizes $A_{ij}$.

The Hamiltonian (\ref{Hamiltonian}) represents an infinite range spin glass with couplings $J_{ij}=(A_{ij}-\gamma p_{ij})$ between all pairs of nodes. The mean of the coupling matrix is zero by construction for $\gamma=1$. Interestingly, we are dealing with a system that shows properties of both the Sherrington-Kirkpatrick (SK) \cite{SK,SKRev} and the Viana-Bray (VB) type \cite{VB}. The distribution of couplings is not symmetric. There are many weak anti-ferromagnetic couplings $J^-$ between unconnected nodes and few but strong ferromagnetic couplings $J^+$ between connected nodes: 
\begin{equation}
J_{ij}^{+}=(1-\gamma p_{ij}),\hspace{0.3cm}J_{ij}^{-}=-\gamma p_{ij},\hspace{0.3cm}J_{ij}^{+}-J_{ij}^{-}=J
\label{Jdef}
\end{equation} 
The parameter $\gamma$ allows tuning the strength of the anti-ferromagnetic interactions and comparing results for different $\gamma$ enables us to detect hierarchies and overlap in community structures \cite{ReichardtPRL, ReichardtPRE}. Note that $J=1$ independent of the choice of the parameter $\gamma$. The individual steps of the following development follow closely along the lines of Fu and Anderson (FA) \cite{FuAnd}, though a treatment along the lines of VB or Monasson \cite{Monasson98} is also possible.

We use the replica trick \cite{EA,SK} in order to evaluate the logarithm of the partition function using $[\ln Z]=\lim_{n\to 0}([Z^n]-1)/n$. The symbol $[\cdot]$ denotes an expectation value over the ensemble of graphs with a given degree distribution. The probability density function of the coupling matrix $J_{ij}$ is then defined as
\begin{equation}
q_{ij}(J_{ij})=p_{ij}\delta(J_{ij}-J_{ij}^{+})+(1-p_{ij})\delta(J_{ij}-J_{ij}^{-}).
\end{equation}
The $p_{ij}$ here is the same as that in equations (\ref{Hamiltonian}) and (\ref{Jdef}). The mean of this distribution is $p(1-\gamma)$ and we are dealing with a diluted ferromagnet for $\gamma<1$ and with a diluted anti-ferromagnet $\gamma>1$. For all $\gamma\geq 1$ we hence expect zero magnetization in the ground state. 

With these preliminaries, we can now follow the steps of FA \cite{FuAnd} and average the replicated partition function over the ensemble of graphs with given degree distribution: 
\begin{equation}
\left[Z^{n}\right]  =  \int \prod_{i,j}dJ_{ij}q_{ij}(J_{ij})\Trn\exp\left\{\beta\sum_{\alpha}^{n}\sum_{i<j}J_{ij}\Sia\Sja\right\}.
\label{Zn2}
\end{equation}
Each of the integrals can be performed separately and after some manipulation we arrive at 
\begin{widetext}
\begin{equation}
\left[Z^n\right]=\Trn\exp\left[\sum_{i<j}\beta\left(J_{ij}^{-}+p_{ij}J\right)\sum_{\alpha}\Sia\Sja+\frac{(\beta J)^2}{2}p_{ij}(1-p_{ij})\left(\sum_{\alpha}^{n}\Sia\Sja\right)^2\right],
\end{equation}
which is still in complete correspondence to Ref. \cite{FuAnd}. Now we deviate from FA's development. Instead of a constant $p_{ij}=p$, we use the factorized form introduced earlier. Neglecting terms involving $p_{ij}^2$ over those involving $p_{ij}$, dropping all terms which vanish for large $N$ or for $n\to0$ and rearranging the sums we arrive at:  
\begin{equation}
\left[Z^n\right]   =  \exp\left(\frac{(\beta J)^2n\langle k\rangle N}{4}\right)\Trn\exp\left\{\frac{\beta (J-\gamma)}{\langle k\rangle N}\sum_{\alpha}\underbrace{\left(\sum_i k_i\Sia\right)^2}_{m_\alpha^{*}}+\frac{(\beta J)^2}{2\langle k\rangle N}\sum_{\alpha<\beta}\underbrace{\left(\sum_i k_i\Sia\Sib\right)^2}_{q_{\alpha\beta}^{*}}\right\}.
\end{equation}
\end{widetext}
Note that we would have arrived at the same expression by using a Gaussian coupling distribution for $q_{ij}(J_{ij})$ with mean $(J-\gamma)/\langle k\rangle N>$ and variance $\sigma^2_{ij}=k_ik_j/\langle k\rangle N$. 
At this point, we have a form that is in principle equivalent to the SK model except for the $k_i$ in the sums over spins. We stress again that setting $\gamma=1$ leads to vanishing magnetization and thus corresponds to a bi-partition. The further development now proceeds along the lines of SK and we arrive at one single site problem for each degree. 

Finally we can derive the replica symmetric equations of state similar to Ref. \cite{Rodgers05}:
\begin{eqnarray}
\label{Mag}
m^* & = & \frac{1}{\langle k\rangle}\sum_k p(k) k\int Dz \tanh(\beta\tilde{H}_k(z)),\\
q^* & = &\frac{1}{\langle k\rangle} \sum_k p(k) k\int Dz\tanh^2(\beta\tilde{H}_k(z)).
\label{SGOrder}
\end{eqnarray} 
The Gaussian measure $Dz=\exp(-z^2/2)/\sqrt{2\pi}$ indicates that the local fields for the spins are also Gaussian distributed, but slightly different for every degree. Nodes of the same degree see the same field distribution. The effective Hamiltonian is
\begin{equation}
\tilde{H}_k(z)=J\sqrt{k q^*}z+(J-\gamma)km^*.
\end{equation}
We can now again follow standard procedures and find the zero temperature limit of the free energy and hence $Q_2$:  
\begin{equation}
Q_2=-\frac{1}{\langle k\rangle} \lim_{\beta\to\infty}[f]=U_0J\langle k^{1/2}\rangle/\langle k\rangle
\label{NewApprox}
\end{equation} 
where $U_0$ is the absolute value of the ground state energy per spin of the Sherrington-Kirkpatrick model. It is interesting to note the similarity with FA, who give $Q_2=U_0\sqrt{(1-p)/Np}$ from the ground state energy of the bi-partitioning problem on Erd\H{o}s-R\'{e}nyi (ER) graphs \cite{Erdos,Bollobas} with link probability $p_{ij}=p$. For graphs with fixed degree, the two formulations of FA and (\ref{NewApprox}) are equivalent. The following numerical experiments will show the adequacy of (\ref{NewApprox}) for a wide range of degree distributions.  

 First ER networks of $N=10,000$ nodes and average degree between $3$ and $20$ were created. The second ensemble of networks is that of scale free networks with $N=10,000$ nodes and a degree distribution of the form $p(k)\propto k^{-\kappa}$ (SF$_{\mbox{\scriptsize kmin}}$). 
The maximum possible degree was set to $1000$ and the graphs were composed using the Molloy-Reed algorithm \cite{MolloyReed}. The second moment of the distribution exists for all $\kappa>2$. We chose $\kappa=3$ for our experiments. In order to produce graphs of different average degree, we introduced a minimum degree $k_{min}$, such that $p(k<k_{min})=0$ with $2\leq k_{min}\leq 12$. Finally, we study a second class of scale free networks (SF$_{\Delta\mbox{k}}$). Introducing a $k_{min}$ may have been a too drastic step, as it excludes all nodes of smaller degree from the network. Therefore, we modify the degree distribution to $p(k)=(k+\Delta k)^{-\kappa}$. Using $\kappa=3$ as before and varying $\Delta k$ between $2$ and $20$. 

\begin{figure}
\includegraphics[width=8cm]{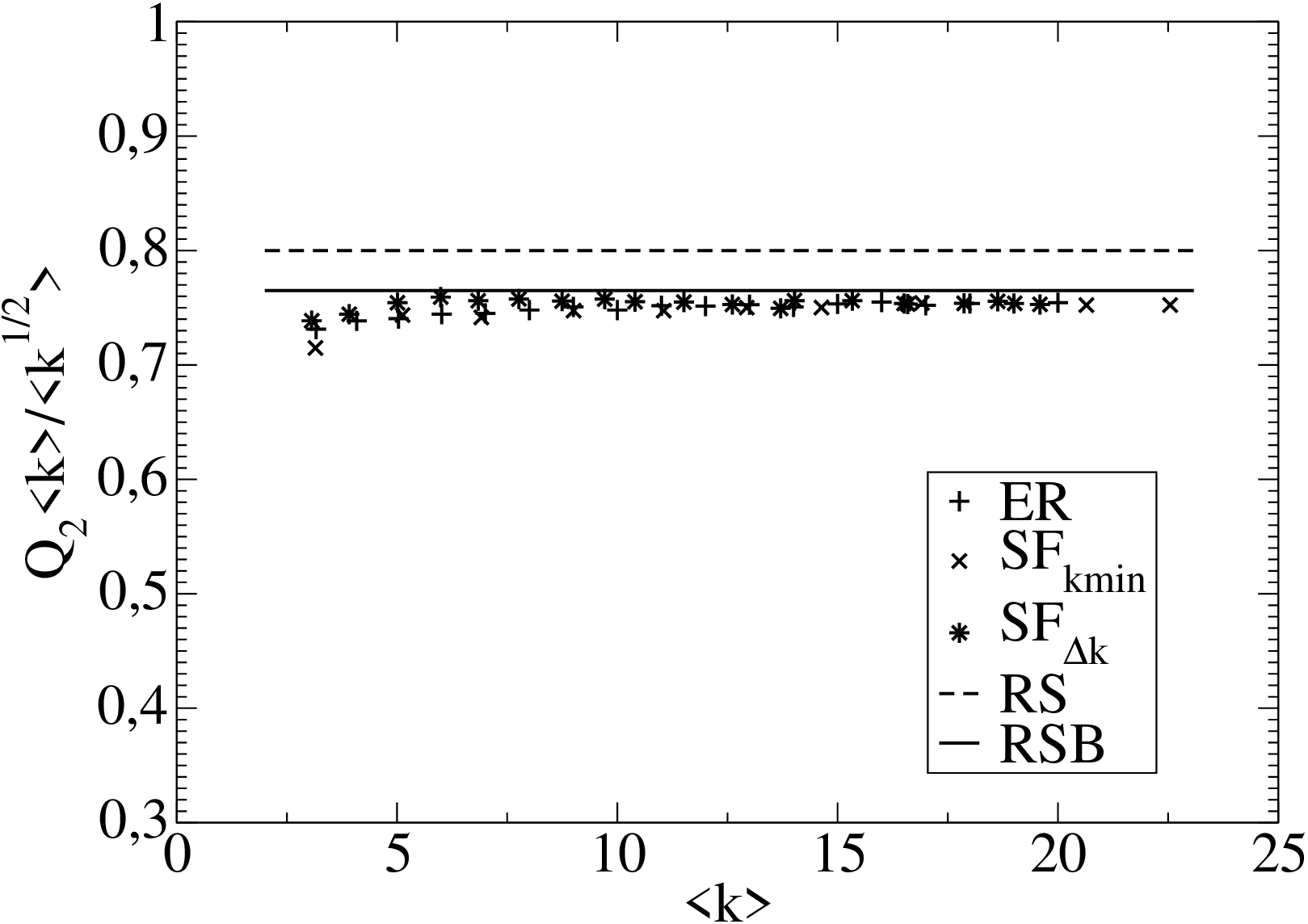}
\caption{Numerical experiments maximizing the modularity of bi-partions in random graphs of different topologies (Erd\H{o}s-R\'{e}nyi random graphs and two forms of a scale free degree distribution, see text for details. Scaling $Q$ in units of $\langle\sqrt{k}\rangle/\langle k\rangle$ collapses the data points onto one universal curve as expected from (\ref{NewApprox}). The dashed and solid lines correspond to $U_0$ in the replica symmetric and replica symmetry breaking case.} 
\label{Q2_rescaled}
\end{figure}

Bi-partitions were calculated by minimizing the Hamiltonian (\ref{Hamiltonian}) using simulated annealing. As expected, the ground state was found to have zero magnetization. Figure \ref{Q2_rescaled} shows the result of this experiment. The empirically found maximum modularity for a bi-partition in networks from these three classes is rescaled in units of $\langle\sqrt{k}\rangle/\langle k\rangle$ and plotted against $\langle k\rangle$ for better visibility. According to (\ref{NewApprox}) we expect the curves to coincide on a horizontal line at $U_0$. Indeed, this rescaling collapses the values for different topologies together. The dashed line represents the ground state energy in the replica symmetric approximation $U_0^{RS}=\sqrt{2/\pi}$ while the solid line corresponds to $U_0^{RSB}=0.765$ \cite{Parisi80,GoldDom90}, the full RSB SK ground state energy. Note that our treatment of the problem allows for the same RSB calculations as the ordinary SK model as all $k_i$ can be drawn out of sums over replica indices. Clearly, we see that (\ref{NewApprox}) gives a good approximation of the modularity and hence cut-size for the different topologies. As expected from the mean field nature of our calculation, the estimate is better for denser graphs. 

\begin{figure}
\includegraphics[width=8cm]{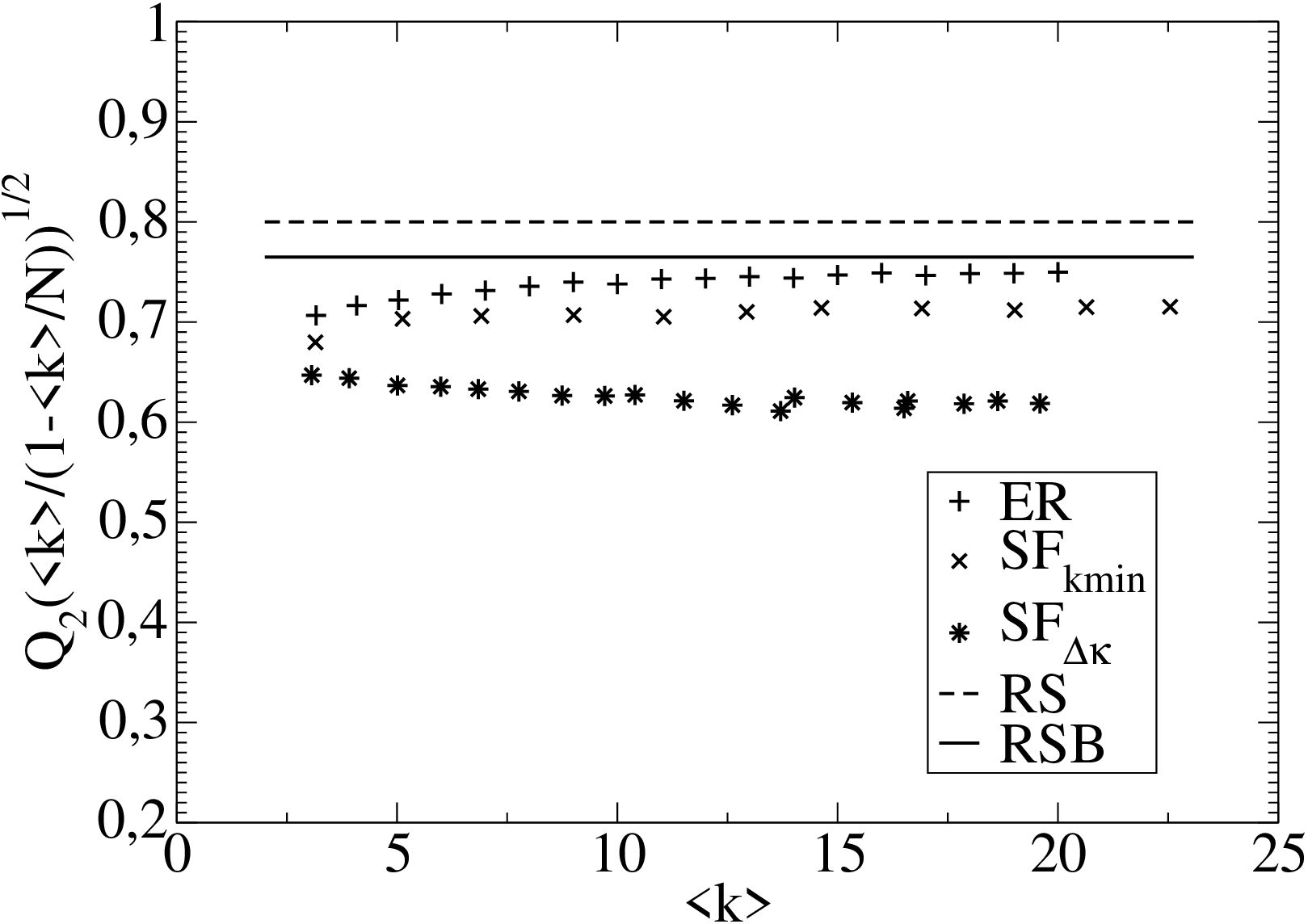}
\caption{The same data points as in Figure \ref{Q2_rescaled}, but $Q$ in units of $\sqrt{(1-p)/Np}$ as suggested by Fu and Anderson \cite{FuAnd}. Clearly, the parametrization by mean degree alone cannot account for the observed cut sizes/modularities in the three different  topologies.}
\label{Q2_rescaled_FA}
\end{figure}

In contrast, Figure \ref{Q2_rescaled_FA} shows the same data points as Figure \ref{Q2_rescaled}, but this time rescaled in units of $\sqrt{(1-p)/\langle k\rangle}$ as suggested by FA. From the fact that the data points do not collapse onto a single universal curve, we can conclude the inadequacy of this scaling for graphs with a broad degree distribution. Note, that (\ref{NewApprox}) also improves over the results of FA in the case of ER graphs. 

\begin{figure}
\includegraphics[width=8cm]{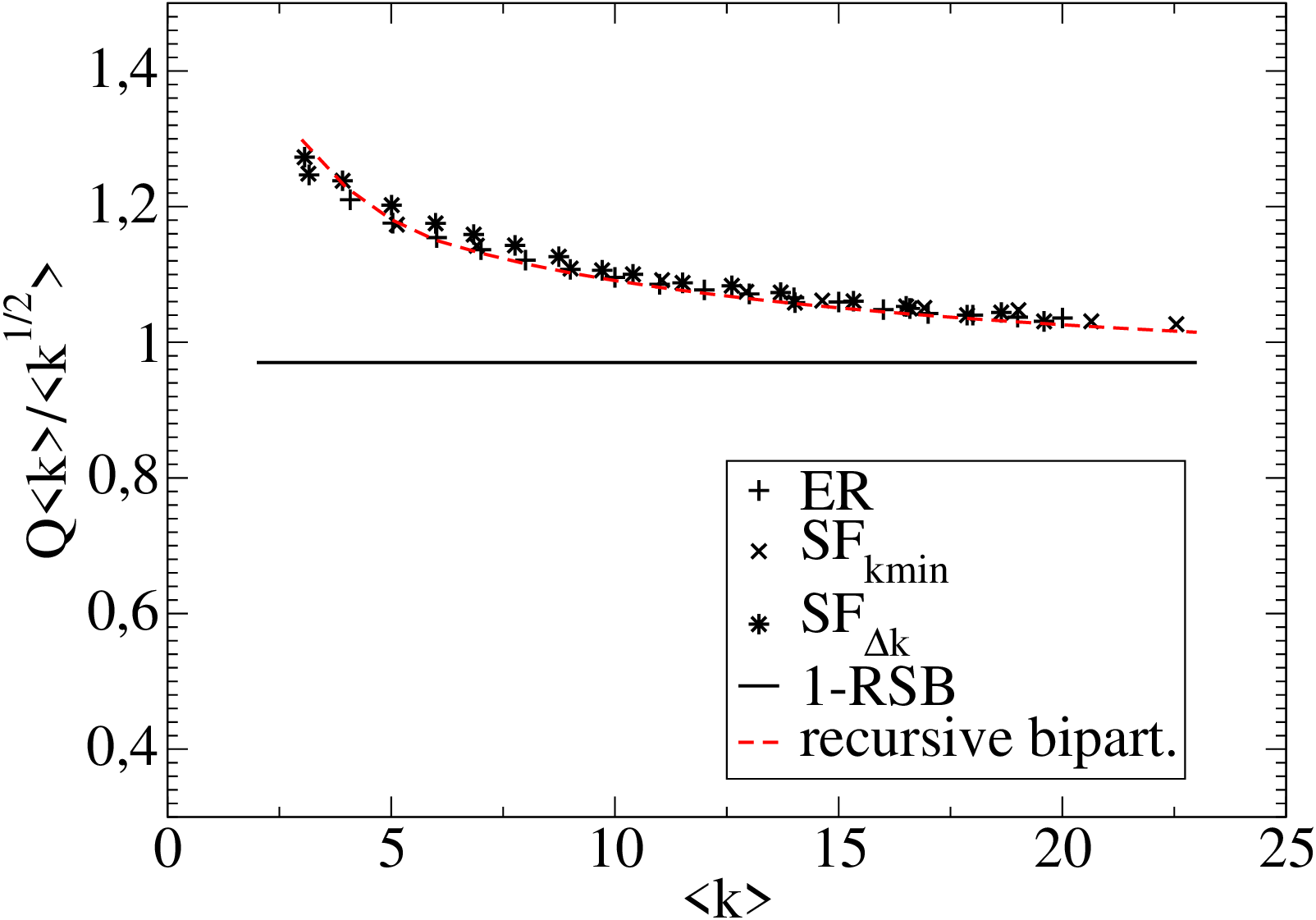}
\caption{Numerical experiments maximizing the modularity in random graphs of different topologies (Erd\H{o}s-R\'{e}nyi random graphs and two forms of a scale free degree distribution, see text for details. Scaling $Q$ in units of $\langle\sqrt{k}\rangle/\langle k\rangle$ collapses the data points onto one universal curve as expected from (\ref{NewApprox}). The solid and solid lines correspond to ground state energy of the Potts glass in 1-RSB approximation due to Kanter and Sompolinsky \cite{KanterSompQPart}. The dashed line corresponds to an estimate obtained from a recursive bi-paritioning along the lines of Ref. \cite{ReichardtPRE} using equation \ref{NewApprox}.}
\label{Q_rescaled}
\end{figure}

Our results from the case of bi-partitioning generalize in a straightforward way 
to the case of q-partitioning and we can simply replace the scaling from the formulas by Kanter and Sompolinsky (KS) \cite{KanterSompQPart} or Lai and Goldschmidt \cite{LaiGold}. The expectation value for the maximum modularity of a random graph with arbitrary degree distribution is then \cite{ReichardtPRE}:
\begin{equation}
Q=0.97\frac{\langle\sqrt k\rangle}{\langle k\rangle}
\label{PottsApprox}
\end{equation}
The factor of $U_0=0.97$ corresponds to the ground state energy of the Potts glass in the 1-RSB treatment as calculated by KS \cite{KanterSompQPart,ReichardtPRE}. Figure \ref{Q_rescaled} shows the maximum modularity obtained when minimizing the Hamiltonian (\ref{Hamiltonian}) in the same graphs as used in Fig. \ref{Q2_rescaled} but with the number of communities $q$ as a free parameter. Again, we see that plotting $Q$ in units of $\langle \sqrt{k}\rangle/\langle k\rangle$ collapses the data points onto a single universal curve. The approximation of the universal value of $U_0$ however, is much slower. This was already observed in Ref. \cite{ReichardtPRE} and can be explained by the fact that sparser graphs tend to cluster into more modules than predicted by KS. This leads to a higher number of degrees of freedom which then tend to accommodate better for fluctuations in the link structure of sparse graphs which lead to relatively higher modularities or lower cut-sizes than expected for denser graphs. Since the rescaling of the modularity has made the data points from all topologies collapse onto a single universal curve, we can employ a recursive estimate using eq. (\ref{NewApprox}) several times as outlined in Ref. \cite{ReichardtPRL}. It is interesting to note that once we have collapsed the data onto a universal curve, we can use the recursive estimate from the convenient ER graphs to give analytical bounds for other topologies as well. 
 
In conclusion, we have presented new approximations for the cut-size of a bi-partition and the corresponding modularity in dense graphs with arbitrary degree distributions. Our results show that scale free networks exhibit larger cut-sizes and smaller modularities than ER graphs of the same average degree. Graphs with a modified scale free degree distribution $p(k)\propto(k+\Delta k)^\kappa$ have larger cut-sizes and smaller modularities than those with a minimum degree. This result also gives new insight into the stability of networks to targeted attack (link removal) \cite{ErrorAttack}. The estimation of modularity of random graphs presented allows for a convenient comparison of the output of community detection algorithms with appropriate random null models and hence gives a first clue on the statistical significance of the findings in real world networks and paves the way to understanding the theoretical limits of community detection \cite{ReichardtPhysicaD}.      

The authors would like to thank Stefan Braunewell, Andreas Engel, Michele Leone, Andrea de Martino, Tobias Galla and Geoff Rodgers, Wolfgang Kinzel, Peter Young, John Hertz and in particular Lenka Zdeborov\'a for many useful hints and discussions. 
\bibliography{../../BibTex_Citations}

\begin{thebibliography}{40}
\expandafter\ifx\csname natexlab\endcsname\relax\def\natexlab#1{#1}\fi
\expandafter\ifx\csname bibnamefont\endcsname\relax
  \def\bibnamefont#1{#1}\fi
\expandafter\ifx\csname bibfnamefont\endcsname\relax
  \def\bibfnamefont#1{#1}\fi
\expandafter\ifx\csname citenamefont\endcsname\relax
  \def\citenamefont#1{#1}\fi
\expandafter\ifx\csname url\endcsname\relax
  \def\url#1{\texttt{#1}}\fi
\expandafter\ifx\csname urlprefix\endcsname\relax\def\urlprefix{URL }\fi
\providecommand{\bibinfo}[2]{#2}
\providecommand{\eprint}[2][]{\url{#2}}

\bibitem[{\citenamefont{Han et~al.}(1998)\citenamefont{Han, Karypis, Kumar, and
  Mobasher}}]{HyperCluster}
\bibinfo{author}{\bibfnamefont{E.-H.} \bibnamefont{Han}},
  \bibinfo{author}{\bibfnamefont{G.}~\bibnamefont{Karypis}},
  \bibinfo{author}{\bibfnamefont{V.}~\bibnamefont{Kumar}}, \bibnamefont{and}
  \bibinfo{author}{\bibfnamefont{B.}~\bibnamefont{Mobasher}},
  \bibinfo{journal}{Data Engineering Bulletin} \textbf{\bibinfo{volume}{21}},
  \bibinfo{pages}{15} (\bibinfo{year}{1998}).

\bibitem[{\citenamefont{Gareya and Johnson}(1979)}]{Garey}
\bibinfo{author}{\bibfnamefont{M.~R.} \bibnamefont{Gareya}} \bibnamefont{and}
  \bibinfo{author}{\bibfnamefont{D.~S.} \bibnamefont{Johnson}},
  \emph{\bibinfo{title}{Computers and Intractability}}
  (\bibinfo{publisher}{W.H.Freeman \& Co Ltd}, \bibinfo{year}{1979}).

\bibitem[{\citenamefont{Erd\H{o}s and R\'{e}nyi}(1960)}]{Erdos}
\bibinfo{author}{\bibfnamefont{P.}~\bibnamefont{Erd\H{o}s}} \bibnamefont{and}
  \bibinfo{author}{\bibfnamefont{A.}~\bibnamefont{R\'{e}nyi}},
  \bibinfo{journal}{Publ. Math. Inst. Hung. Acad. Sci.}
  \textbf{\bibinfo{volume}{5}}, \bibinfo{pages}{17} (\bibinfo{year}{1960}).

\bibitem[{\citenamefont{Bollobas}(2001)}]{Bollobas}
\bibinfo{author}{\bibfnamefont{B.}~\bibnamefont{Bollobas}},
  \emph{\bibinfo{title}{Random Graphs}} (\bibinfo{publisher}{Cambridge
  University Press}, \bibinfo{year}{2001}), \bibinfo{edition}{2nd} ed.

\bibitem[{\citenamefont{Fu and Anderson}(1986)}]{FuAnd}
\bibinfo{author}{\bibfnamefont{Y.}~\bibnamefont{Fu}} \bibnamefont{and}
  \bibinfo{author}{\bibfnamefont{P.~W.} \bibnamefont{Anderson}},
  \bibinfo{journal}{J. Phys. A: Math. Gen.} \textbf{\bibinfo{volume}{19}},
  \bibinfo{pages}{1605} (\bibinfo{year}{1986}).

\bibitem[{\citenamefont{Kanter and Sompolinsky}(1987)}]{KanterSompQPart}
\bibinfo{author}{\bibfnamefont{I.}~\bibnamefont{Kanter}} \bibnamefont{and}
  \bibinfo{author}{\bibfnamefont{H.}~\bibnamefont{Sompolinsky}},
  \bibinfo{journal}{J. Phys. A: Math. Gen.} \textbf{\bibinfo{volume}{20}},
  \bibinfo{pages}{L636} (\bibinfo{year}{1987}).

\bibitem[{\citenamefont{Lai and Goldschmidt}(1987)}]{LaiGold}
\bibinfo{author}{\bibfnamefont{P.-Y.} \bibnamefont{Lai}} \bibnamefont{and}
  \bibinfo{author}{\bibfnamefont{Y.~Y.} \bibnamefont{Goldschmidt}},
  \bibinfo{journal}{J. Stat. Phys.} \textbf{\bibinfo{volume}{48}},
  \bibinfo{pages}{513} (\bibinfo{year}{1987}).

\bibitem[{\citenamefont{Liao}(1988)}]{LiaoRev}
\bibinfo{author}{\bibfnamefont{W.}~\bibnamefont{Liao}}, \bibinfo{journal}{Phys.
  Rev. A} \textbf{\bibinfo{volume}{37}}, \bibinfo{pages}{587}
  (\bibinfo{year}{1988}).

\bibitem[{\citenamefont{Liao}(1987)}]{LiaoPRL}
\bibinfo{author}{\bibfnamefont{W.}~\bibnamefont{Liao}}, \bibinfo{journal}{Phys.
  Rev. Lett.} \textbf{\bibinfo{volume}{59}}, \bibinfo{pages}{1625}
  (\bibinfo{year}{1987}).

\bibitem[{\citenamefont{Goldschmidt and Dominicis}(1990)}]{GoldDom90}
\bibinfo{author}{\bibfnamefont{Y.~Y.} \bibnamefont{Goldschmidt}}
  \bibnamefont{and} \bibinfo{author}{\bibfnamefont{C.~D.}
  \bibnamefont{Dominicis}}, \bibinfo{journal}{Phys. Rev. B}
  \textbf{\bibinfo{volume}{410}}, \bibinfo{pages}{2184} (\bibinfo{year}{1990}).

\bibitem[{\citenamefont{Wong and Sherrington}(1987)}]{Wong}
\bibinfo{author}{\bibfnamefont{K.~Y.~M.} \bibnamefont{Wong}} \bibnamefont{and}
  \bibinfo{author}{\bibfnamefont{D.}~\bibnamefont{Sherrington}},
  \bibinfo{journal}{J. Phys. A: Math. Gen.} \textbf{\bibinfo{volume}{20}},
  \bibinfo{pages}{L793} (\bibinfo{year}{1987}).

\bibitem[{\citenamefont{Mezard and Parisi}(1987)}]{Mezard87}
\bibinfo{author}{\bibfnamefont{M.}~\bibnamefont{Mezard}} \bibnamefont{and}
  \bibinfo{author}{\bibfnamefont{G.}~\bibnamefont{Parisi}},
  \bibinfo{journal}{Europhys. Lett.} \textbf{\bibinfo{volume}{3}},
  \bibinfo{pages}{1067} (\bibinfo{year}{1987}).

\bibitem[{\citenamefont{Banavar et~al.}(1987)\citenamefont{Banavar,
  Sherrington, and Sourlas}}]{Banavar}
\bibinfo{author}{\bibfnamefont{J.~R.} \bibnamefont{Banavar}},
  \bibinfo{author}{\bibfnamefont{D.}~\bibnamefont{Sherrington}},
  \bibnamefont{and} \bibinfo{author}{\bibfnamefont{N.}~\bibnamefont{Sourlas}},
  \bibinfo{journal}{J. Phys. A: Math. Gen.} \textbf{\bibinfo{volume}{20}},
  \bibinfo{pages}{L1} (\bibinfo{year}{1987}).

\bibitem[{\citenamefont{de~Oliveira}(1989)}]{Oliveira89}
\bibinfo{author}{\bibfnamefont{M.~J.} \bibnamefont{de~Oliveira}},
  \bibinfo{journal}{J. Stat. Phys.} \textbf{\bibinfo{volume}{54}},
  \bibinfo{pages}{477} (\bibinfo{year}{1989}).

\bibitem[{\citenamefont{Sherrington and Wong}(1987)}]{SherringtonBethe}
\bibinfo{author}{\bibfnamefont{D.}~\bibnamefont{Sherrington}} \bibnamefont{and}
  \bibinfo{author}{\bibfnamefont{K.~Y.~M.} \bibnamefont{Wong}},
  \bibinfo{journal}{J. Phys. A: Math. Gen.} \textbf{\bibinfo{volume}{20}},
  \bibinfo{pages}{L785} (\bibinfo{year}{1987}).

\bibitem[{\citenamefont{Steinbach et~al.}(2003)\citenamefont{Steinbach,
  Ert\"oz, and Kumar}}]{ClusterChallenges}
\bibinfo{author}{\bibfnamefont{M.}~\bibnamefont{Steinbach}},
  \bibinfo{author}{\bibfnamefont{L.}~\bibnamefont{Ert\"oz}}, \bibnamefont{and}
  \bibinfo{author}{\bibfnamefont{V.}~\bibnamefont{Kumar}},
  \emph{\bibinfo{title}{New Vistas in Statistical Physics -- Applications in
  Econo-physics, Bioinformatics, and Pattern Recognition}}
  (\bibinfo{publisher}{Springer-Verlag}, \bibinfo{year}{2003}), chap.
  \bibinfo{chapter}{Challenges of clustering high dimensional data}.

\bibitem[{\citenamefont{Cheng et~al.}(2001)\citenamefont{Cheng, Kahng, and
  Liu}}]{VLSIPowerLaw}
\bibinfo{author}{\bibfnamefont{C.-K.} \bibnamefont{Cheng}},
  \bibinfo{author}{\bibfnamefont{A.~B.} \bibnamefont{Kahng}}, \bibnamefont{and}
  \bibinfo{author}{\bibfnamefont{B.}~\bibnamefont{Liu}}, in
  \emph{\bibinfo{booktitle}{SLIP '01: Proceedings of the 2001 international
  workshop on System-level interconnect prediction}} (\bibinfo{publisher}{ACM
  Press}, \bibinfo{address}{New York, NY, USA}, \bibinfo{year}{2001}), pp.
  \bibinfo{pages}{99--106}.

\bibitem[{\citenamefont{Guimera and Amaral}(2005)}]{GuimeraMeta}
\bibinfo{author}{\bibfnamefont{R.}~\bibnamefont{Guimera}} \bibnamefont{and}
  \bibinfo{author}{\bibfnamefont{L.~A.~N.} \bibnamefont{Amaral}},
  \bibinfo{journal}{Nature} \textbf{\bibinfo{volume}{433}},
  \bibinfo{pages}{895} (\bibinfo{year}{2005}).

\bibitem[{\citenamefont{Palla et~al.}(2005)\citenamefont{Palla, Derenyi,
  Farkas, and Vicsek}}]{Palla05}
\bibinfo{author}{\bibfnamefont{G.}~\bibnamefont{Palla}},
  \bibinfo{author}{\bibfnamefont{I.}~\bibnamefont{Derenyi}},
  \bibinfo{author}{\bibfnamefont{I.}~\bibnamefont{Farkas}}, \bibnamefont{and}
  \bibinfo{author}{\bibfnamefont{T.}~\bibnamefont{Vicsek}},
  \bibinfo{journal}{Nature} \textbf{\bibinfo{volume}{435}},
  \bibinfo{pages}{814} (\bibinfo{year}{2005}).

\bibitem[{\citenamefont{Arenas et~al.}(2004)\citenamefont{Arenas, Diaz-Guilera,
  Gleiser, and Guimera}}]{Arenas}
\bibinfo{author}{\bibfnamefont{A.}~\bibnamefont{Arenas}},
  \bibinfo{author}{\bibfnamefont{A.}~\bibnamefont{Diaz-Guilera}},
  \bibinfo{author}{\bibfnamefont{P.}~\bibnamefont{Gleiser}}, \bibnamefont{and}
  \bibinfo{author}{\bibfnamefont{R.}~\bibnamefont{Guimera}},
  \bibinfo{journal}{Eur. Phys. J. B} \textbf{\bibinfo{volume}{38}},
  \bibinfo{pages}{373} (\bibinfo{year}{2004}).

\bibitem[{\citenamefont{Newman and Girvan}(2003)}]{Girvan}
\bibinfo{author}{\bibfnamefont{M.~E.~J.} \bibnamefont{Newman}}
  \bibnamefont{and} \bibinfo{author}{\bibfnamefont{M.}~\bibnamefont{Girvan}},
  \bibinfo{journal}{Proc. Natl. Acad. Sci. USA} \textbf{\bibinfo{volume}{99}},
  \bibinfo{pages}{7821} (\bibinfo{year}{2003}).

\bibitem[{\citenamefont{Danon et~al.}(2005)\citenamefont{Danon, Dutch, Arenas,
  and Diaz-Guilera}}]{GuileraReview}
\bibinfo{author}{\bibfnamefont{L.}~\bibnamefont{Danon}},
  \bibinfo{author}{\bibfnamefont{J.}~\bibnamefont{Dutch}},
  \bibinfo{author}{\bibfnamefont{A.}~\bibnamefont{Arenas}}, \bibnamefont{and}
  \bibinfo{author}{\bibfnamefont{A.}~\bibnamefont{Diaz-Guilera}},
  \bibinfo{journal}{J. Stat. Mech.} p. \bibinfo{pages}{P09008}
  (\bibinfo{year}{2005}).

\bibitem[{\citenamefont{Newman}(2003)}]{NewmanMixing}
\bibinfo{author}{\bibfnamefont{M.~E.~J.} \bibnamefont{Newman}},
  \bibinfo{journal}{Phys. Rev. E.} \textbf{\bibinfo{volume}{67}},
  \bibinfo{pages}{026126} (\bibinfo{year}{2003}).

\bibitem[{\citenamefont{Newman and Girvan}(2004)}]{Girvan03}
\bibinfo{author}{\bibfnamefont{M.~E.~J.} \bibnamefont{Newman}}
  \bibnamefont{and} \bibinfo{author}{\bibfnamefont{M.}~\bibnamefont{Girvan}},
  \bibinfo{journal}{Phys. Rev. E} \textbf{\bibinfo{volume}{69}},
  \bibinfo{pages}{026113} (\bibinfo{year}{2004}).

\bibitem[{\citenamefont{Guimera et~al.}(2004)\citenamefont{Guimera,
  Sales-Pardo, and Amaral}}]{Guimera}
\bibinfo{author}{\bibfnamefont{R.}~\bibnamefont{Guimera}},
  \bibinfo{author}{\bibfnamefont{M.}~\bibnamefont{Sales-Pardo}},
  \bibnamefont{and} \bibinfo{author}{\bibfnamefont{L.~N.}
  \bibnamefont{Amaral}}, \bibinfo{journal}{Phys. Rev. E.}
  \textbf{\bibinfo{volume}{70}}, \bibinfo{pages}{025101(R)}
  (\bibinfo{year}{2004}).

\bibitem[{\citenamefont{Reichardt and
  Bornholdt}(2006{\natexlab{a}})}]{ReichardtPRE}
\bibinfo{author}{\bibfnamefont{J.}~\bibnamefont{Reichardt}} \bibnamefont{and}
  \bibinfo{author}{\bibfnamefont{S.}~\bibnamefont{Bornholdt}},
  \bibinfo{journal}{Phys. Rev. E} \textbf{\bibinfo{volume}{74}},
  \bibinfo{pages}{016110} (\bibinfo{year}{2006}{\natexlab{a}}).

\bibitem[{\citenamefont{Engel and den Broeck}(2001)}]{EngelBuch}
\bibinfo{author}{\bibfnamefont{A.}~\bibnamefont{Engel}} \bibnamefont{and}
  \bibinfo{author}{\bibfnamefont{C.~V.} \bibnamefont{den Broeck}},
  \emph{\bibinfo{title}{Statistical Mechanics of Learning}}
  (\bibinfo{publisher}{Cambridge University Press}, \bibinfo{year}{2001}).

\bibitem[{\citenamefont{Newman}(2004)}]{FastGN}
\bibinfo{author}{\bibfnamefont{M.~E.~J.} \bibnamefont{Newman}},
  \bibinfo{journal}{Phys. Rev. E.} \textbf{\bibinfo{volume}{69}},
  \bibinfo{pages}{066133} (\bibinfo{year}{2004}).

\bibitem[{\citenamefont{Reichardt and Bornholdt}(2004)}]{ReichardtPRL}
\bibinfo{author}{\bibfnamefont{J.}~\bibnamefont{Reichardt}} \bibnamefont{and}
  \bibinfo{author}{\bibfnamefont{S.}~\bibnamefont{Bornholdt}},
  \bibinfo{journal}{Phys. Rev. Lett.} \textbf{\bibinfo{volume}{93}},
  \bibinfo{pages}{218701} (\bibinfo{year}{2004}).

\bibitem[{\citenamefont{Chung and Lu}(2002)}]{ChungLu}
\bibinfo{author}{\bibfnamefont{F.}~\bibnamefont{Chung}} \bibnamefont{and}
  \bibinfo{author}{\bibfnamefont{L.}~\bibnamefont{Lu}},
  \bibinfo{journal}{Annals of Combinatorics} \textbf{\bibinfo{volume}{6}},
  \bibinfo{pages}{125} (\bibinfo{year}{2002}).

\bibitem[{\citenamefont{Sherrington and Kirkpatrick}(1975)}]{SK}
\bibinfo{author}{\bibfnamefont{D.}~\bibnamefont{Sherrington}} \bibnamefont{and}
  \bibinfo{author}{\bibfnamefont{S.}~\bibnamefont{Kirkpatrick}},
  \bibinfo{journal}{Phys. Rev. Lett.} \textbf{\bibinfo{volume}{35}},
  \bibinfo{pages}{1792} (\bibinfo{year}{1975}).

\bibitem[{\citenamefont{Kirkpatrick and Sherrington}(1977)}]{SKRev}
\bibinfo{author}{\bibfnamefont{S.}~\bibnamefont{Kirkpatrick}} \bibnamefont{and}
  \bibinfo{author}{\bibfnamefont{D.}~\bibnamefont{Sherrington}},
  \bibinfo{journal}{Phys. Rev. B} \textbf{\bibinfo{volume}{17}}
  (\bibinfo{year}{1977}).

\bibitem[{\citenamefont{Viana and Bray}(1985)}]{VB}
\bibinfo{author}{\bibfnamefont{L.}~\bibnamefont{Viana}} \bibnamefont{and}
  \bibinfo{author}{\bibfnamefont{A.~J.} \bibnamefont{Bray}},
  \bibinfo{journal}{J. Phys. C: Solid State Phys.}
  \textbf{\bibinfo{volume}{18}}, \bibinfo{pages}{3037} (\bibinfo{year}{1985}).

\bibitem[{\citenamefont{Monasson}(1998)}]{Monasson98}
\bibinfo{author}{\bibfnamefont{R.}~\bibnamefont{Monasson}},
  \bibinfo{journal}{J. Phys. A: Math. Gen.} \textbf{\bibinfo{volume}{31}},
  \bibinfo{pages}{513} (\bibinfo{year}{1998}).

\bibitem[{\citenamefont{Edwards and Anderson}(1975)}]{EA}
\bibinfo{author}{\bibfnamefont{S.~F.} \bibnamefont{Edwards}} \bibnamefont{and}
  \bibinfo{author}{\bibfnamefont{P.~W.} \bibnamefont{Anderson}},
  \bibinfo{journal}{J. Phys. F. Met. Phys} \textbf{\bibinfo{volume}{5}},
  \bibinfo{pages}{965} (\bibinfo{year}{1975}).

\bibitem[{\citenamefont{Kim et~al.}(2005)\citenamefont{Kim, Rodgers, Kahng, ,
  and Kim}}]{Rodgers05}
\bibinfo{author}{\bibfnamefont{D.-H.} \bibnamefont{Kim}},
  \bibinfo{author}{\bibfnamefont{G.~J.} \bibnamefont{Rodgers}},
  \bibinfo{author}{\bibfnamefont{B.}~\bibnamefont{Kahng}}, , \bibnamefont{and}
  \bibinfo{author}{\bibfnamefont{D.}~\bibnamefont{Kim}},
  \bibinfo{journal}{Phys. Rev. E} \textbf{\bibinfo{volume}{71}},
  \bibinfo{pages}{056115} (\bibinfo{year}{2005}).

\bibitem[{\citenamefont{Molloy and Reed}(1995)}]{MolloyReed}
\bibinfo{author}{\bibfnamefont{M.}~\bibnamefont{Molloy}} \bibnamefont{and}
  \bibinfo{author}{\bibfnamefont{B.}~\bibnamefont{Reed}},
  \bibinfo{journal}{Random Structures and Algorithms}
  \textbf{\bibinfo{volume}{6}}, \bibinfo{pages}{161} (\bibinfo{year}{1995}).

\bibitem[{\citenamefont{Parisi}(1980)}]{Parisi80}
\bibinfo{author}{\bibfnamefont{G.}~\bibnamefont{Parisi}}, \bibinfo{journal}{J.
  Phys. A: Math. Gen.} \textbf{\bibinfo{volume}{13}}, \bibinfo{pages}{L115}
  (\bibinfo{year}{1980}).

\bibitem[{\citenamefont{Albert et~al.}(2000)\citenamefont{Albert, Jeong, and
  Barab\'{a}si}}]{ErrorAttack}
\bibinfo{author}{\bibfnamefont{R.}~\bibnamefont{Albert}},
  \bibinfo{author}{\bibfnamefont{H.}~\bibnamefont{Jeong}}, \bibnamefont{and}
  \bibinfo{author}{\bibfnamefont{A.-L.} \bibnamefont{Barab\'{a}si}},
  \bibinfo{journal}{Nature} \textbf{\bibinfo{volume}{406}},
  \bibinfo{pages}{378} (\bibinfo{year}{2000}).

\bibitem[{\citenamefont{Reichardt and
  Bornholdt}(2006{\natexlab{b}})}]{ReichardtPhysicaD}
\bibinfo{author}{\bibfnamefont{J.}~\bibnamefont{Reichardt}} \bibnamefont{and}
  \bibinfo{author}{\bibfnamefont{S.}~\bibnamefont{Bornholdt}},
  \bibinfo{journal}{Physica D} \textbf{\bibinfo{volume}{224}},
  \bibinfo{pages}{20} (\bibinfo{year}{2006}{\natexlab{b}}).

\end{thebibliography}

\end{document}